\documentclass[11pt]{article}

\title{Evidence and Strategy on Economic Distance in Spatially Augmented Solow-Swan Growth Model}
\author{Jieun Lee \\  Department of Economics, \\ University of Illinois Urbana-Champaign, USA \\  \textit{jieun2@illinois.edu}}

\usepackage{inputenc} 

\usepackage{amsmath}
\usepackage{tabularx}
\usepackage{adjustbox}
\usepackage{amsthm}
\usepackage{mathtools}
\usepackage{undertilde}
\usepackage{amssymb}
\usepackage{mathrsfs} 
\usepackage{setspace}
\usepackage{multirow}
\usepackage{caption}
\usepackage{subcaption}
\usepackage{graphics}
\usepackage{blindtext}
\usepackage{rotating}

\usepackage[english]{babel} 
\usepackage{amssymb}
\usepackage{amsmath}
\usepackage{mathdots}
\usepackage{breqn}
\usepackage{graphicx}
\usepackage[figurename=Figure]{caption}
\usepackage[labelsep=endash]{caption}
\newcounter{Figcount}
\newcounter{tempFigure}

\usepackage{amsfonts}
\usepackage{setspace}

\theoremstyle{definition}
\newtheorem{proposition}{Proposition}

\makeatletter
\renewcommand{\@seccntformat}[1]{%
	\ifcsname prefix@#1\endcsname
	\csname prefix@#1\endcsname
	\else
	\csname the#1\endcsname\quad
	\fi}

\makeatother

\begin{document}
	\onehalfspacing
	
	\date{}
	\maketitle
	\vspace{-5mm}
	\begin{abstract}
		\indent Economists' interests in growth theory have a very long history (Harrod, 1939; Domar, 1946; Solow, 1956; Swan 1956; Mankiw, Romer, and Weil, 1992). Recently, starting from the neoclassical growth model, Ertur and Koch (2007) developed the spatially augmented Solow-Swan growth model with the exogenous spatial weights matrices ($W$). While the exogenous $W$ assumption could be true only with the geographical/physical distance, it may not be true when economic/social distances play a role. Using Penn World Table version 7.1, which covers year 1960-2010, I conducted the robust Rao's score test (Bera, Dogan, and Taspinar, 2018) to determine if $W$ is endogeonus and used the maximum likelihood estimation (Qu and Lee, 2015). The key finding is that the significance and positive effects of physical capital externalities and spatial externalities (technological interdependence) in Ertur and Koch (2007) were no longer found with the exogenous $W$, but still they were with the endogenous $W$ models. I also found an empirical strategy on which economic distance to use when the data recently has been under heavy shocks of the worldwide financial crises during year 1996-2010. 
		
		
		
	\end{abstract}
	
	\newpage
	\section{Introduction} 
	\noindent Exploring economic growth and convergence theories helps us to better understand the source of gaps between rich and poor countries. A provocative question can be raised, whether all countries have the equal (or a single absolute) or conditional convergence equilibrium by different covariates. The history of the neoclassical growth model began by challenging Harrod (1939) and Domar (1946) growth model with unstable steady-state growth. Solow (1956), Swan (1956), and Meade (1961) asserted that the capital-output ratio is not exogenous but regresses back the economy to its steady-state growth path. In particular, the total output in this model depends on the two factors--accumulated stock of physical capital and the working-age population within a Cobb-Douglas production function and a determined steady-state output per capita. Extensions of the basic Solow-Swan model by adding other socio-economic covariates have been suggested by Barro (1991), Mankiw, Romer, and Weil (1992, henceforth MRW), and Barro and Sala-i-Martin (2004). Notably, MRW (1992) made an additional extension by decomposing the capital as physical and human capital, following Lucas (1988, 1993)'s argument for human capital externalities as a determinant of economic growth.  \\
	
	Solow-Swan growth model and its various extensions did not take into account of the cross-country externalities. In the literature, however, there is plenty of evidence about spillover effects across countries or cross-country externalities: for instance, see De Long and Summers, 1991; Temple, 1999; Keller, 2002; Klenow and Rodriguez-Clare, 2005. Driven by these empirical evidence, Ertur and Koch (2007) presented an augmented MRW model with the knowledge effects accumulated within home country \& its spatial externalities to neighborhood and found the significantly positive spatial externalities across countries at steady-state. The spatial weights matrices ($W$) in Ertur and Koch (2007) were given geographical, i.e. \textit{exogenous}. The knowledge spillover effects through economic interactions, however, have been found empirically significant: For example, see literature on international trade (Frankel and Rose, 1998; Baxter and Kouparitas, 2005; Ditzen, 2018). In this spirit, the bilateral trade flow is used as $W$ (Ertur and Koch, 2011; Ho, Wang, and Yu, 2013). Also, Table 1 shows a list of top ten trading countries with the US according to United States Census Bureau\footnote{Year-to-Date Total Trade, reported in April 2021. https://www.census.gov/foreign-trade/statistics/highlights/toppartners.html.} and the estimated flight time with average airplane speed of 567mph\footnote{Distance calculator, GlobeFeed.com. https://distancecalculator.globefeed.com/Country_Distance_Calculator.asp}. United Kingdom, for example, is the 7th trading country with the US, while they rank the 3rd shortest rank in flight time. This implies the physical proximity does not necessarily lead to more economic interactions. Hence spatial interactions could be more generalized to depend on both the economic variables as well as the predetermined geography. One may take various forms of physical/economic distances featuring their characteristics. \\
	
	\begin{table}[h!]
		\centering
		\caption{Trade volume rank vs. Shortest flight time rank}
		\small
		\begin{tabular}{cc|cc} \hline
			Trade Rank & Trade (\%)             & Estimated Flight Time & Shortest Rank \\ \hline
			1          & Mexico (14.8\%)        & 1.79                  & 1             \\
			2          & Canada (14.4\%)        & 2.48                  & 2             \\
			3          & China (14.0\%)         & 12.76                 & 7             \\
			4          & Japan (4.8\%)          & 11.12                 & 5             \\
			5          & Germany (4.5\%)        & 8.61                  & 4             \\
			6          & Korea, South (3.6\%)   & 11.77                 & 6             \\
			7          & United Kingdom (2.6\%) & 7.49                  & 3             \\
			8          & Vietnam (2.5\%)        & 15.12                 & 10            \\
			9          & India(2.4\%)           & 14.88                 & 9             \\
			10         & Taiwan (2.4\%)         & 13.41                 & 8            \\ \hline
		\end{tabular}
	\end{table}
	
	Considering the fact that economic variables are closely related to each other, however, such $W$ should \textit{not} be treated exogenous since $W$ may not be exogenous but \textit{stochastic} and   \textit{endogenous} if the variables in $W$ and the dependent variable are correlated (Qu and Lee, 2015). Accordingly, one needs to test if $W$ is endogenous. For this, the robust Rao's score test (Bera, Dogan, and Taspinar, 2018) is used since it is computationally efficient and valid with respect to the type 1 error,  irrespective of a presence in local misspecification as will be discussed in Section 3.2. The source of endogeneity and the maximum likelihood estimator with the endogenous $W$ (Qu and Lee, 2015) will be discussed in Section 3.1.\\
	
	The rest of the article is organized as follows. In section 2, the spatially-augmented Solow-Swan growth model in Ertur and Koch (2007) is reviewed. Section 3 introduces the source of endogeneity in $W$ and the maximum likelihood estimators (Qu and Lee, 2015) and the robust Rao's score test (Bera, Dogan, and Taspinar, 2018) to determine if $W$ is endogenous. In section 4, the data and variables as well as a couple of physical/economic distance forms are described. In section 5, the results are presented and the key findings are discussed. In section 6, I conclude. 
	
	\section{Spatially augmented Solow-Swan Growth model}
	Driven by earlier literature which emphasized the technological spillover effects and cross-country externalities, Ertur and Koch (2007) introduced the spatially augmented growth model by adding spatially weighted neighborhood values of the dependent variable and covariates. Specifically, the production function for a country \textit{i} $(i=1, ..., N)$ takes Cobb-Douglas form but with time index \textit{t} ($t=1,...,T$): 
	\begin{equation} \label{GrindEQ__2_3_1_} 
	Y_i\left(t\right)=A_i\left(t\right)K^{\alpha }_i\left(t\right)L^{1-\alpha }_i\left(t\right), 
	\end{equation} 
	\noindent where $Y_i(t)$ is the output, $K_i(t)$ is the level of reproducible physical capital, ${\mathrm{L}}_{\mathrm{i}}(t)$ is the level of labor. Specification differs from the traditional neoclassical growth model by explicitly modeling $A_i(t)$ instead of assuming it a constant scale of the level of the technology; More specifically, $A_i(t)$ is the aggregate level of technology with
	\begin{equation} \label{GrindEQ__2_3_2_} 
	A_i\left(t\right)=\ \mathit{\Omega}\left(t\right)k^{\phi }_i\left(t\right)\prod^N_{j\neq i}{A^{\gamma w_{ij}}_j\left(t\right)},                                      
	\end{equation} 
	\noindent where $\mathit{\Omega}\left(t\right)=\mathit{\Omega}\left(0\right)e^{\mu t}$ represents some proportion of technological progress as exogenous and identical in all countries, with its constant rate of growth \textit{$\mu$}; $\textit{k}_i(t)=\frac{K_i(t)}{L_i(t)}$ is the accumulated physical capital per capita; \textit{$\phi$} ($0\leq \phi < 1$) denotes the connectivity of physical capital externalities or home externalities with respect to knowledge from capital investment; $\gamma$ ($0 \leq \gamma <1$) represents the technological interdependence across countries or neighbors. Finally,  $w_{ij}$ represents the connectivity between a country \textit{i} and \textit{j} with non-negative, non-stochastic and finite with $0\le w_{ij}<1$\textit{, }$w_{ij}=0$\textit{ }if\textit{ }$i=j,$  and row-normalized so that $\sum\limits_{j\neq i}^N{w_{ij}=1}$ for $\mathrm{i=1,\ \dots ,\ N.}$ Combining equations (1) and (2), we have 
	\begin{equation} \label{GrindEQ__2_3_3_} 
	y_i\left(t\right)={\mathit{\Omega}}^{\frac{1}{1-\gamma }}\left(t\right)k^{u_{ii}}_i(t)\prod^N_{j\neq i}{k^{u_{ij}}_j\left(t\right),}\  
	\end{equation} 
	\noindent with $u_{ii}=\alpha +\phi (1+\sum\limits_{r=1}^\infty{{\gamma }^rw^{\left(r\right)}_{ii}})$ and $u_{ij}=\phi \sum\limits_{r=1}^\infty{{\gamma }^rw^{\left(r\right)}_{ij}}.$ $u_{ii}$ represents country \textit{i}'s social return if country \textit{i } increases its own stock of physical capital per worker. Thus the aggregate social return for the country \textit{i} would be $u_{ii} + \sum\limits_{j \neq i}^N u_{ij}=\alpha+\frac{\phi}{1-\gamma}$ if all countries concurrently increase their stocks of physical capital per worker. The terms $w^{\left(r\right)}_{ij}$ are the elements of row \textit{i} and column \textit{j} of W to the power of r and hence represent the second-, third-, and higher order spatial connections. $y_i(t)={Y_i(t)}/{L_i(t)}$ is the level of output per worker. Using equation (2) at steady-state $k_i^*=\Omega^{\frac{1}{(1-\gamma)(1-u_{ii})}}(t)\left(\frac{s_i}{n_i+g+\delta}\right)^{\frac{1}{1-u_{ii}}}\prod\limits_{j \neq i}^N k_j^{*\frac{u_{ij}}{1-u_{ii}}}(t)$  gives the following spatial regression equation as
	
	\begin{align}
	lny_i&=\beta_0+\beta_1lns_i+\beta_2ln(n_i+g+\delta)\\
	&\hspace{3mm}+\theta_1\sum_{j \neq i}^N w_{ij}lns_j+\theta_2\sum_{j\neq i}^Nw_{ij}ln(n_j+g+\delta)+\rho\sum_{j \neq i}^N w_{ij}lny_i+\epsilon_i,
	\end{align}
	
	\noindent where $s_i$ is a constant fraction of output saved and $n_i$ is the exogenous labor growth rate for a country \textit{i}. $\delta$ is the annual rate of depreciation of physical capital for all countries and assumed constant; $g$ is the balanced growth rate with $g=\frac{\mu}{(1-\alpha)(1-\gamma)-\phi}$. Here $\frac{1}{1-\alpha-\phi}ln \Omega(0)=\beta_0+\epsilon_i $ and $(g+\delta)$ was taken as 0.05 in Mankiw, Romer, and Weil (MRW, 1992) and Romer (1989). The theoretical constraints to be tested are $\theta_2=-\theta_1=\frac{\alpha\gamma}{1-\alpha-\phi}$ and $\beta_1=-\beta_2=\frac{\alpha+\phi}{1-\alpha-\phi}$. Equation (4) can be rewritten in the matrix form of spatial Durbin model (SDM) (Durbin, 1960):
	\begin{equation}
	Y=\rho WY+X_2\beta+WX_1\theta+\epsilon,
	\end{equation}
	\noindent where $Y=(lny_i)_{i=1,...,n}$, $X_1=(lns_j, ln(n_j+g+\delta))_{i=1,...,n}$, \\$X_2=(constant, lns_i,ln(n_i+g+\delta))_{i=1,...,n}$, $\theta=(\theta_1, \theta_2)'$ and $\beta=(\beta_0, \beta_1, \beta_2)'.$ Further, it could be rewritten as the form of spatial autoregressive model (SAR):
	\begin{equation}
	Y=\rho WY+X\beta+\epsilon,
	\end{equation}
	\noindent where $X=(constant,lns_i,ln(n_i+g+\delta),\sum\limits_{j \neq i}^N w_{ij}\mathrm{ln}s_j, \sum\limits_{j \neq i}^N w_{ij}\mathrm{ln}(n_j+g+\delta))_{i=1,...,n}$ and $\beta=(\beta_0, \beta_1, \beta_2, \theta_1, \theta_2)'.$ 
	
	\section{Spatial Weights Matrices on the Economic Distance}
	\subsection{ML specification}
	Following Jenish and Prucha (2009, 2012), Qu and Lee (2015) adopted the asymptotic inference under near-epoch dependence (NED). $W_n$ is allowed to be asymmetric. Specifically, the main regression model is 
	\begin{equation}
	Y_n=\rho W_nY_n + X_{1n}\beta + V_n,
	\end{equation}
	\noindent where $Y_n$ is an $n \times 1$ vector of dependent variable; $X_{1n}$ is an $n \times k_1$ matrix of regressors, $\beta$ is a $k_1 \times 1$ vector of coefficients; $V_n$ is an $n \times 1$ vector of the disturbance term in the main regression; and $W_n$ is an n×n non-negative matrix with zero diagonals. The elements $w_{ij,n}$ of $W_n$  are constructed by $Z_n$ and $\rho_{ij}$: $w_{ij,n}=h_{ij{}(Z_n,\rho_{ij})}$ for $i,j=1,…,n \hspace{2mm} \forall i \neq j, \hspace{2mm}$ where $\rho_{ij} (\geq 1$ without loss of generality) is the physical distance between country \textit{i} and \textit{j} and $h(\cdot)$ is a bounded function. \\
	
	The auxiliary equation for $Z_n$ is
	\begin{equation}
	Z_n=X_{2n}\Gamma+\varepsilon_n,
	\end{equation}
	\noindent where $Z_n$ is an $n\times p$ matrix; $X_{2n}$ is an $n \times k_2$ matrix; $\Gamma$ is $k_2 \times 1$ vector of coefficients; and $\varepsilon_n$ is an $n \times 1$ vector of the error term in equation (23). Furthermore, the error terms $v_{i,n}$ and $\varepsilon_{i,n}$ are assumed to have the joint normal distribution: 
	
	\begin{equation}
	(v_{i,n},\varepsilon_{i,n}') \overset{\textit{iid}}{\sim}  N(0,\Sigma_{v \varepsilon}),
	\end{equation}
	
	\noindent where $\Sigma_{v \varepsilon}=\begin{pmatrix} \sigma_v^2 & \sigma_{v \varepsilon}' \\ \sigma_{v\varepsilon} & \Sigma_\varepsilon \end{pmatrix}$ is positive definite; $\sigma_v^2$ is a scalar variance, $\sigma_{v\varepsilon}$ is a $p$ dimensional vector; and $\Sigma_\varepsilon$ is a $p \times p$ matrix. Letting 
	\begin{equation}
	\delta=\Sigma_\varepsilon^{-1}\sigma_{v\varepsilon}
	\end{equation}
	\noindent and $\sigma_\xi^2=\sigma_v^2-\sigma_{v\varepsilon}'\Sigma_\varepsilon^{-1}\sigma_{v\varepsilon}$, the conditional mean and variance are rewritten as $E(\nu_{i,n} \vert \varepsilon_{i,n} )=\varepsilon_{i,n}'\delta, Var(\nu_{i,n}\vert\varepsilon_{i,n} )=\sigma_\xi^2.$ Now define $\xi_n$ by 
	\begin{equation}
	\xi_n=V_n-\varepsilon_n\delta.
	\end{equation}
	Then $\xi_n$ has conditional mean zero (on $\varepsilon_n$) and conditional variance of $\sigma_\xi^2 I_n.$ \\
	
	The source of endogeneity is found if the covariance of $(\nu_{i,n},\varepsilon_{i,n}')$ is nonzero or $\sigma_{\nu\varepsilon} \neq 0$. To control the endogeneity, the equations (7) and (8) are integrated, which yields
	\begin{equation}
	Y_n=\rho W_nY_n + X_{1n}\beta + (Z_n-X_{2n}\Gamma)\delta+\xi_n,
	\end{equation}
	\noindent with $E(\xi_{i,n} \vert \varepsilon_{i,n})=0$, $E(\xi_{i,n}^2 \vert \varepsilon_{i,n})=\sigma_\xi^2$, and $\xi_{i,n}'s$ \textit{iid}. That is, $Z_n-X_{2n}\Gamma$ is used as the control variable to deal with the endogeneity of $W_n$. The log likelihood function is then
	
	\begin{align*}
	lnL_n &=-nln(2\pi)-\frac{n}{2}ln|\Sigma_{\nu\varepsilon}|+ln|S_n(\rho)| \\
	&\hspace{3mm}-\frac{1}{2}[(S_n(\rho)Y_n-X_{1n}\beta),(vec(Z_n-X_{2n}\Gamma))']\\
	&\hspace{3mm}\times(\Sigma_{\nu\varepsilon}^{-1}\otimes I_n)\begin{pmatrix} S_n(\rho)Y_n-X_{1n}\beta \\
	vec(Z_n-X_{2n}\Gamma) \end{pmatrix},
	\end{align*}
	
	\noindent where $S_n(\rho)=I_n-\rho W_n$. Letting $\mu=(\rho,\beta',vec(\Gamma)',\sigma_\xi^2,\alpha',\delta')'$ gives the log likelihood function as
	
	\begin{align*}
	lnL_n(\mu)&=-nln(2\pi)-\frac{n}{2}\sigma_\xi^2+ln|S_n(\rho)|-\frac{n}{2}ln|\Sigma_\varepsilon|\\
	&\hspace{3mm} -\frac{1}{2}\sum\limits_{i=1}^n (z_{i,n}'-x_{2,in}'\Gamma)\Sigma_\varepsilon^{-1}(z_{i,n}-\Gamma'X_{2,in})\\
	&\hspace{3mm} -\frac{1}{2\sigma_\xi^2}[S_n(\rho)Y_n-X_{1n}\beta-(Z_n-X_{2n}\Gamma)\delta]'\\
	&\hspace{3mm} \times [S_n(\rho)Y_n-X_{1n}\beta-(Z_n-X_{2n}\Gamma)\delta].
	\end{align*}

	\begin{proposition}
		Under regular assumptions, the maximum likelihood estimator $\hat{\mu}$ is a consistent estimator of $\theta_0$ and
		\begin{equation*}
		\sqrt{n}(\hat{\mu}-\mu_0)\xrightarrow{d}N\left(0,\lim\limits_{n \rightarrow \infty}\frac{1}{n}E\left(\frac{\partial^2 lnL_L(\mu_0)}{\partial \mu \partial \mu'}\right)^{-1}\right).
		\end{equation*}
	\end{proposition}
	\textbf{Proof.} See Theorem 3 in Qu and Lee (2015).
	
	\subsection{Testing Endogeneity}
	Note that one should test if $W$ is endogenous, i.e., if one rejcts $H_0^\delta: \delta=0$ before regular analysis. Specifically, recall that $\mu=(\rho,\beta',vec(\Gamma)',\sigma_\xi^2,\alpha',\delta')'$ in the previous subsection. Consider the following sequences of local alternatives:
	\begin{align*}
	&H_A^\delta: \delta_0=\Delta_\delta/\sqrt{n} \\
	&H_A^\rho: \rho_0=\Delta_\rho/\sqrt{n}, 
	\end{align*}
	
	\noindent where $\Delta_\delta$ and $\Delta_\rho$ are bounded vectors. Under the joint null ($H_0^\delta$ and $H_0^\rho$), the restricted ML estimator $\tilde{\mu}$ is defined by $(0,\beta',vec(\Gamma)',\sigma_\xi^2,\alpha',0')'$. We may actually regard $\tilde{\mu}=(0,\beta',0')'$ since the information matrix at $\tilde{\mu}$ is block diagonal with respect to $vec(\Gamma),\sigma_\xi^2$ and $\alpha$. Let $L_\psi(\mu)=\frac{1}{n}\frac{\partial lnL(\mu)}{\partial \psi}$ and $I(\mu)=-\frac{1}{n}\frac{\partial^2 lnL(\mu)}{\partial \mu \partial \mu'}$, where $\psi \in \{\rho,\beta,\delta\}$. The standard langrange multiplier (LM) test statistic is defined as
	\begin{equation*}
	LM_\delta(\tilde{\mu})=nL_\delta'(\tilde{\mu})I_{\delta \cdot \beta}(\tilde{\mu})^{-1}L_\delta(\tilde{\mu}).
	\end{equation*}
	
	\begin{proposition}
		Under regular assumptions, the following results hold.
		
		1. Under $H_A^\delta$ and $H_A^\rho$,
		\begin{equation*}
		LM_\delta(\tilde{\mu})\xrightarrow{d}\chi^2_{p}(\varphi_1),
		\end{equation*}
		where $\varphi_1=\Delta_\delta'I_{\delta\cdot\beta}(\mu_0)\Delta_\delta+\Delta_\delta'I_{\delta\rho\cdot\beta}(\mu_0)\Delta_\rho+\Delta_\rho'I_{\delta\rho\cdot\gamma}'(\mu_0)\Delta_\delta$\\$+\Delta_\rho'I_{\delta\rho\cdot\beta}'(\mu_0)I_{\delta\cdot\beta}^{-1}(\mu_0)I_{\delta\rho\cdot\beta}(\mu_0)\Delta_\rho$ is the non-centrality parameter. \\
		
		2. Under $H_0^\delta: \delta_0=0$ and irrespective of whether $H_0^\rho$ or $H_A^\rho$ holds, the distribution of the robust test $LM_\delta^*(\tilde{\mu})$ is
		\begin{equation*}
		LM_\delta^*(\tilde{\mu})=nL_\delta^*(\tilde{\mu})[I_{\delta\cdot\beta}(\tilde{\mu})-I_{\delta\rho\cdot \beta}(\tilde{\mu})I_{\rho \cdot \beta}^{-1}(\tilde{\mu})I_{\delta\rho\cdot\beta}'(\tilde{\mu})]^{-1}L_\delta^*(\tilde{\mu}) \xrightarrow{d} \chi^2_{p},
		\end{equation*}
		where $L_\delta^*(\tilde{\mu})=[L_\delta(\tilde{\mu})-I_{\delta\rho \cdot \beta}(\tilde{\mu})I_{\rho \cdot \beta}^{-1}(\tilde{\mu})L_\lambda(\tilde{\mu})]$ is the adjusted score function, $I_{\delta\rho \cdot \beta}(\tilde{\mu})=I_{\delta\rho}(\tilde{\mu})-I_{\delta\beta}(\tilde{\mu})I_{\beta\beta}^{-1}(\tilde{\mu})I_{\beta\rho}(\tilde{\mu})$ and $I_{\rho \cdot \beta}(\tilde{\mu})=I_{\rho\rho}(\tilde{\mu})-I_{\rho\beta}(\tilde{\mu})I_{\beta\beta}^{-1}(\tilde{\mu})I_{\beta\rho}(\tilde{\mu})$. \\
		
		3. Under $H_A^\delta$ and irrespective of whether $H_0^\rho$ or $H_A^\rho$ holds,
		\begin{equation*}
		LM_\delta^*(\tilde{\mu}) \xrightarrow{d} \chi^2_{p}(\varphi_2),
		\end{equation*}
		where $\varphi_2=\Delta_\delta'(I_{\delta\cdot \beta}(\mu_0)-I_{\delta\rho \cdot \beta}(\mu_0)I_{\rho\cdot\beta}^{-1}(\mu_0)I_{\delta\rho\cdot\beta}'(\mu_0)\Delta_\delta$ is the non-centrality parameter.
	\end{proposition}
	\textbf{Proof.} See Proposition 1 in Bera, Dogan, and Taspinar (2018). \\
	
	\noindent Remark that the robust Rao's score test is computationally efficient over the conditional LM tests (Qu and Lee, 2015; Cheng and Lee, 2017) in that it only requires restricted estimators under $H_0^\rho: \rho_0=0$. RS test is also robust in the sense that its limiting distribution is asymptotically central chi-squared regardless of local misspecification in $\rho_0$, where the standard RS test yields over-rejection of $H_0^\delta$.  
	
	\section{Data}
	
	\noindent The data is extended from Penn World Table version 6.1 (Ertur and Koch, 2007) to version 7.1 and the year is also extended from 1960-1995 to 1960-2010\footnote{There are many missing values before year 1960 and hence omitted.}. Ninety countries of the MRW (1992) non-oil sample is used and listed in Table 2. Note that the periods of the worldwide financial crisis (late 1990s and 2000s) are quite recent, wherefore the shocks are still effective in this dataset. The dependent and explanatory variables as well as the measure of each variable are exactly the same as in Ertur and Koch (2007). The number of workers is computed by RGDPCH $\times$ POP / RGDPW, where RGDPCH is real GDP per capita by the chain method, RGDPW is real-chain GDP per worker, and POP is the total population. Real income per worker is measured by RGDPW. The saving rate $s$ is measured from the share of gross investment (KI) in GDP. \\ 
	
	\begin{table}[]
		\caption{List of countries}
		\tiny
		\centering
		\addtolength{\leftskip} {-2cm}
		\addtolength{\rightskip}{-2cm}
		\begin{tabular}{c|c|llllllllllll} \hline
			Letter & Numbers & \multicolumn{12}{l}{Countries}                                                                                                                \\ \hline
			A      & 3       & \multicolumn{12}{l}{Argentina, Australia, Austria,}                                                                                           \\
			B      & 8       & \multicolumn{12}{l}{Bangladesh, Belgium, Benin, Bolivia, Botswana,   Brazil, Burkina Faso, Burundi}                                           \\
			C      & 10      & \multicolumn{12}{l}{Cameroon, Canada, Central African Rep., Chad,   Chile, Colombia, Congo, Dem.Rep., Congo, Rep., Costa Rica, Cote D’Ivoire} \\
			D      & 2       & \multicolumn{12}{l}{Denmark, Dominican Rep.}                                                                                                  \\
			E      & 4       & \multicolumn{12}{l}{Ecuador,   Egypt, El Salvador, Ethiopia}                                                                                  \\
			F      & 2       & \multicolumn{12}{l}{Finland, France}                                                                                                          \\
			G      & 3       & \multicolumn{12}{l}{Ghana, Greece, Guatemala}                                                                                                 \\
			H      & 2       & \multicolumn{12}{l}{Honduras, Hong Kong}                                                                                                      \\
			I      & 5       & \multicolumn{12}{l}{India, Indonesia,    Ireland, Israel, Italy}                                                                              \\
			J      & 3       & \multicolumn{12}{l}{Jamaica, Japan, Jordan}                                                                                                   \\
			K      & 2       & \multicolumn{12}{l}{Kenya, Korea Rep. of}                                                                                                     \\
			M      & 9       & \multicolumn{12}{l}{Madagascar, Malawi, Malaysia, Mali, Mauritania,   Mauritius, Mexico, Morocco, Mozambique}                                 \\
			N      & 7       & \multicolumn{12}{l}{Nepal, Netherlands, New Zealand, Nicaragua, Niger,   Nigeria, Norway}                                                     \\
			P      & 7       & \multicolumn{12}{l}{Pakistan, Papua New Guinea, Panama, Paraguay, Peru,   Philippines, Portugal}                                              \\
			R      & 1       & \multicolumn{12}{l}{Rwanda}                                                                                                                   \\
			S      & 9       & \multicolumn{12}{l}{Senegal, Sierra Leone, Singapore, South Africa,   Spain, Sri Lanka, Sweden, Switzerland, Syria}                           \\
			T      & 6       & \multicolumn{12}{l}{Tanzania, Thailand, Togo, Trinidad Tobago, Tunisia,   Turkey}                                                             \\
			U      & 4       & \multicolumn{12}{l}{Uganda, United Kingdom, Uruguay, United States of   America}                                                              \\
			V      & 1       & \multicolumn{12}{l}{Venezuela}                                                                                                                \\
			Z      & 2       & \multicolumn{12}{l}{Zambia, Zimbabwe}                         \\ \hline                                                                               
		\end{tabular}
	\end{table}
	
	Note that an economic distance measures the similarity or proximity of two economies with respect to a specific economic variable. I further assume diminishing returns to scale of economic variables as a general sense. I thus consider the three types of spatial weights matrices as follows:
	
	\begin{equation*}
	\begin{cases}
	w_{1,ij}^e=exp(-2d_{ij}) \\
	w_{2,ij}^e=1/|z_i-z_j|^{2\cdot s*} \\
	w_{3,ij}^e=1/(min\left(\frac{z_i}{z_j},\frac{z_j}{z_i}\right)-1)^2,\\
	\end{cases}
	\end{equation*}
	
	\noindent for all $i \neq j$ and zeros if $i=j$. $w_{1,ij}$ is a negative exponential function and $s*$ in $w_{2,ij}$ is a tuning parameter which gives the best result with respect to finding the significance of the parameters of interest in Table 3. Note that $w_{2,ij}^e$ and $w_{3,ij}^e$ are of the power distance weights function with different forms of economic similarity: $w_{2,ij}$ gives more weights if the difference in an economic variable is smaller; $w_{3,ij}^e$ gives more weights if the ratio of an economic variable gets closer to 1. They all describe diminishing effects in distance. The 3D plots and their contours of economic distances above are described in Figure 1, obtained from the free software of Wolfram alpha.  \\
	
	\begin{figure}
		\caption{3D plots and their contours}
		\begin{subfigure}{.3\textwidth}
			\centering
			\includegraphics[width=.8\linewidth]{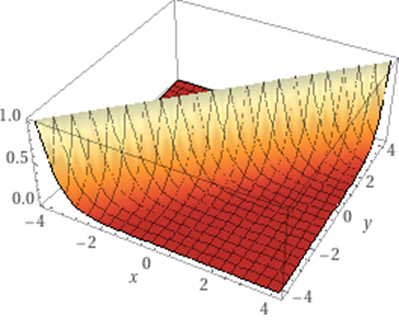}  
			\caption{3D plot of $w^e_{1,ij}$}
			\label{fig:sub-first}
		\end{subfigure}
		\begin{subfigure}{.3\textwidth}
			\centering
			\includegraphics[width=.8\linewidth]{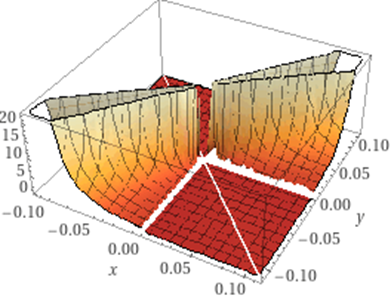}  
			\caption{3D plot of $w^e_{2,ij}$}
			\label{fig:sub-second}
		\end{subfigure}
		\begin{subfigure}{.3\textwidth}
			\centering
			\includegraphics[width=.8\linewidth]{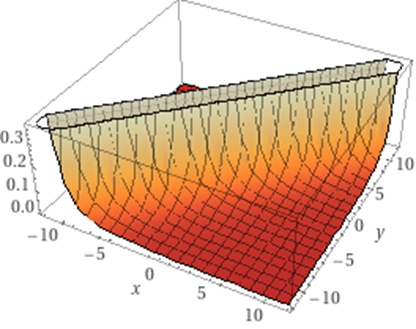}  
			\caption{3D plot of $w^e_{3,ij}$}
			\label{fig:sub-second}
		\end{subfigure}
		\newline
		
		\begin{subfigure}{.3\textwidth}
			\centering
			\includegraphics[width=.8\linewidth]{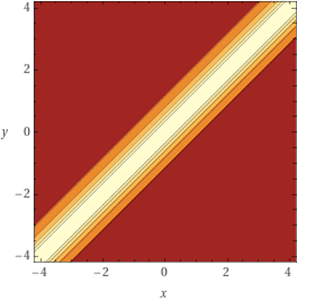}  
			\caption{Contour of $w^e_{1,ij}$}
			\label{fig:sub-third}
		\end{subfigure}
		\begin{subfigure}{.3\textwidth}
			\centering
			\includegraphics[width=.8\linewidth]{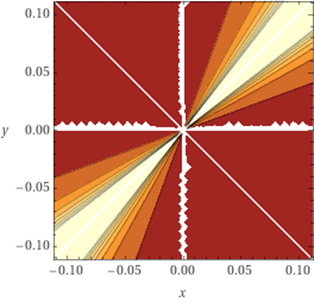}  
			\caption{Contour of $w^e_{2,ij}$}
			\label{fig:sub-fourth}
		\end{subfigure}
		\begin{subfigure}{.3\textwidth}
			\centering
			\includegraphics[width=.8\linewidth]{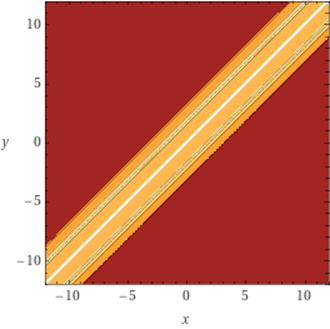}  
			\caption{Contour of $w^e_{3,ij}$}
			\label{fig:sub-second}
		\end{subfigure}
		\label{fig:fig}
	\end{figure}

	For the physical distance, we use the great-circle distance in Ertur and Koch (2007). The great-circle distance is defined by	
	\begin{equation*}
	D_{ij}=radius \times cos^{-1}[cos|long_i-long_j|conslat_i costlat_j+sinlat_i sinlat_j],
	\end{equation*}
	\noindent where \textit{radius} is the Earth’s radius, and \textit{lat} and \textit{long} are latitude and longitude for \textit{i} and \textit{j}. The physical weights matrices given in Ertur and Koch (2007) are
	
	\begin{equation*}
	\begin{cases}
	w^d_{1,ij}=D^{-2}_{ij} \\
	w^d_{2,ij}=exp(-2D_{ij}), \\
	\end{cases}
	\end{equation*}
	
	\noindent if $i \neq j$ and zero if $i=j$. We denote $W^d_{1n}=(w^d_{1,ij})_{i,j=1,\neq,n}$ and $W^d_{2n}=(w^d_{2,ij})_{i,j=1,\neq,n}$. We then obtain Hadamard product of $(w^d_{q_1,ij})_{i,j=1,\dots,n}$ and $(w^e_{q_2,ij})_{i,j=1,\dots,n}$, i.e., $(w^d_{q_1,ij} \circ w^e_{q_2,ij})_{i,j=1,\dots,n}$ for all $q_1=1,2; q_2=1,2,3$ and normalize it. \\
	
	Now we suspect that the real \textit{Gross Domestic Income} (GDI) meaningfully constructs the spatial weights matrices across countries with respect to real income per worker (RGDPW) because GDI measures economic activities based on all kinds of income to manage standard of living. Note that GDI and GDP are theoretically equivalent but not equal in practice (Figure 2, Visualization chart from Federal Reserve Economic Data) since the economy is very complicated with a lot of measurement errors and the components are measured from different sectors. Additionally, empirical evidence assert inequality between GDI and GDP more than just the measurement error: For example, GDI has diagnosed economic downturns better than GDP has (Nalewaik, 2012). We hence take GDI and GDP differently in this paper and expect GDI to construct $W$ pertaining to real income per worker. \\
	
	\begin{figure}
		\centering
		\caption{GDI vs. GDP}
		\includegraphics[scale=0.6]{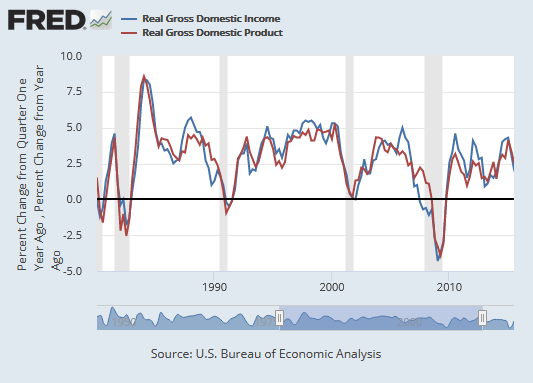}
	\end{figure}
	
	\begin{table}[hbt!]
		\centering
		\caption{\small Summary of $Z_n$ \& $X_{2n}$ and the parameters by the dependent variable}
		\footnotesize
		\begin{tabular}{c|cc} \hline
			\centering
			\multirow{2}{*}{Dependent   variable} & \multicolumn{2}{c}{\multirow{2}{*}{Log of real income per worker ($lny$)}} \\ 
			& &  \\ \hline
			$Z_n$                                     & \multicolumn{2}{c}{Real GDI (RGDPTT)}\\ 
			\multirow{1}{*}{$X_{2n}$}                   & \multicolumn{2}{c}{constant terms, log of $s$   ($lns$)}          \\ \hline
			Parameters of Interest & \multicolumn{2}{c}{Proportion of physical capital ($\alpha$)} \\
			& \multicolumn{2}{c}{Physical capital externalities ($\phi$)} \\
			& \multicolumn{2}{c}{Technological interdependence ($\gamma$)} \\ \hline
		\end{tabular}
	\end{table}

	\section{Results}
	Table 4 and Table 5 present the results. Table 4 covers the period over 1960-2010 using Penn World Table version 7.1 and Table 5 is for year 1960-1995 using the earlier version of 6.1. Both in Table 4 and Table 5, Model (1) is the classical Solow-Swan growth model. Model (2)-(5) and Model (6)-(9) respectively have common physical weights matrices of $W^d_{1n}$ and $W^d_{2n}$, while their economic distance types vary. Model (2) \& (6) have an exogenous $W^d$, which is the model in Ertur and Koch (2007) and Model (3) \& (7) have the exponential form of $w^e_{ij}=exp(-2d_{ij})$, where $d_{ij}$ is the Euclidean distance. Model (4) \& (8) have a form of power distance weights and the tuning parameter, which gives the best result, is found to be 2.5. Model (5) \& (9) compares the farness of the ratio between $z_i$ and $z_j$ from 1 and gives higher weights as they get close to 1. \\
	
	Table 4 shows that all robust RS tests reject the null that $W$ is exogenous and thus one needs the control variables to deal with the endogeneity of $W$. The key finding is that the spatial model with the exogenous $W$ (Model (2)) gives totally contradictory results to those in Ertur and Koch (2007) after 15 years where the worldwide financial crisis occured in late 1990s through 2000s. All parameters of physical capital share ($\alpha$), physical capital externalities ($\phi$), and technological interdependence ($\gamma$) are not  significant and the signs of $\phi,\gamma$ are the opposite depending on the physical weights matrices ($W^d_{1n}$ or $W^d_{2n}$). The presence of $\phi$ is not significant, either. This is interesting because they were highly significant and positive in Ertur and Koch (2007). However, the models with the endogenous $W$ could capture the significances of $\gamma$ and $\phi$ as well as their positive effects (Model (4)-(5), (8)-(9)). The presence of $\phi$ is also reported strongly significant at level 0.01. Moreover, I find a positive externality of the neighbor's saving rate (\textit{s}), while one's saving rate is not significant at all against Model (2): a country $i$'s $lns$ is highly significant if $W$ is exogenous, while $lns$ in the country $i$'s neighborhood is rather highly significant and postive if $W$ is endogenous. The population growth rate in the neighborhood also gives the opposite results against Model (2). Furthermore, the estimates of the spatial autoregressive parameter ($\rho$) are also greater in the endogenous $W$ than those in the exogenous $W$. All these findings indicate that the spatial interactions become very different once the economic distance comes into play. One would deduce that economic distance makes newly generated stance for economies on top of the exogenously given physical distance. \\
	
	I also find some consistent results across the models in Table 4. The theoretical constraints hold in all spatially augmented Solow-Swan models (Model (2)-(5), Test of Restriction). The existence of the physical capital externalities or technological interdependence will not lead to the endogenous growth in the economy since $\alpha+\frac{\phi}{1-\gamma}<1$. Model (4) \& (8) and (5) \& (9)  respectively show more or less comparable results in terms of estimating the effects of regressors. Nevertheless, there is a nonnegligible discrepancy in the magnitude of the estimated parameters ($\phi, \gamma$). Even though their significance was captured across the two different types of $W^d$, their estimates warn us to apply more appropriate physical weight matrix which fits better to the economic variables that construct $W^e$. The way the effect of economic distance spreads out may depend on the feature of the physical distance such as slope, curvature, or any other geometric properties. Additionally, note that the estimates for $\alpha$ are either overestimated (Model (2)) or negative (Model (4)-(5), (8)-(9)) against our expectation but insignificant across all models except for the original Solow-Swan model. The overestimation in Model (2) may lie in the frequent and worldwide financial crisis in late 1990s through 2000s so that the effect of the physical capital is unclear to the real GDP (RGDPCH) as well as the real chain GDP per worker (RGDPWOK), while the endogenous $W$ is specialized in capturing the physical ($\phi$) and spatial externalities ($\gamma$) but weak at capturing the effects of physical capital share ($\alpha$) (see Table 5). \\
	
	Table 5 covers the period over year 1960-1995 using Penn World Table version 6.1, where the world was before experiencing the global financial crises and therefore relatively stable with regard to physical capital shock. The results in Ertur and Koch (2007) are rewritten in Table 4 and our models with endogenous $W$ (Model (3)-(5), (7)-(9)) are presented. Recall that all parameters ($\alpha, \phi, \gamma$) are significant and positive (Model (2)). Interestingly, Model (4) and Model (5) which showed nice results in Table 4 are now failing in capturing significance of any parameters, while Model (3) secures the positive and significant effects of the parameters as Model (2). Model (3) is also robust to the different types of $W^d$. This result may suggest an empirical strategy on which type of economic distance to use: if the period is relatively stable with respect to physical capital, an exogenous $W$ is good enough as well as an endogenous $W$ whose economic distance is of the negative exponential function, while an endogenous $W$ with the power distance weight function works better otherwise. A detailed analysis on the different properties of power law and negative exponential law in the distance-decay function is found in Chen (2015) as regard to the geographcial gravity model.

	\begin{sidewaystable}
		\caption{Results using Penn World Table verison 7.1 (Year 1960-2010)}
		\begin{minipage}{0.5\textwidth}
			\tiny
			\begin{tabular}{c|c|cccc|cccc}  \cline{2-10}
				&                                                                              & & \multicolumn{3}{c}{$W^d_{1n} \circ W^e_n$}  \vline                                                                                                                                                                                                                     & \multicolumn{4}{c}{$W^d_{2n} \circ W^e_n$}                                                                                                                                                                                                                       
				\\
				& Solow-Swan                                                                        &  \multicolumn{1}{c}{$W^d_{1n}$}                                                   & $exp(-2d_{ij})$                                                       & $\frac{1}{(|z_i-z_j|^{2 \cdot s*})}$                                    & $\frac{1}{(min\left(\frac{z_i}{z_j},\frac{z_j}{z_i}\right)-1)^2}$                                                     &        $W^d_{2n}$                                               & $exp(-2d_{ij})$                                                      & $\frac{1}{(|z_i-z_j|^{2 \cdot s*})}$                                       & $\frac{1}{(min\left(\frac{z_i}{z_j},\frac{z_j}{z_i}\right)-1)^2}$                                                     \\ \hline
				Model & (1) & (2) & (3) & (4) & (5) & (6) & (7) & (8) & (9) \\ \hline
				Endogeneity test (RS)  & -                                                                            & -                                                         & 44.726***                                                 & 24.9675***                                                & 18.515***                                                 & -                                                         & 41.591***                                                 & 29.126***                                                 & 29.112***                                                 \\
				Constant terms                      & -0.238                                                                       & 1.183                                                     & 1.319***                                                  & -1.281***                                                 & -0.327                                                    & 0.702                                                     & 0.758                                                     & -0.258                                                    & 1.207***                                                  \\
				$lns$                       & 1.928***                                                                     & 1.256***                                                  & 1.244                                                     & 0.474                                                     & 0.592                                                     & 1.284***                                                  & 1.289                                                     & 0.422                                                     & 0.619                                                     \\
				$lnngd$                      & -4.837***                                                                    & -2.729***                                                 & -2.738***                                                 & -0.704***                                                 & -0.622***                                                 & -2.478***                                                 & -2.500***                                                 & -0.745***                                                 & -0.684***                                                 \\
				$Wlns$                      & -                                                                            & 0.789*                                                    & 0.721***                                                  & 0.077***                                                  & 0.031***                                                  & 0.214                                                     & 0.252***                                                  & 0.311***                                                  & 0.325***                                                  \\
				$Wln(n+g+\delta)$                    & -                                                                            & 0.247                                                     & 0.420***                                                  & -1.065***                                                 & -0.787***                                                 & 0.107                                                     & 0.084                                                     & -0.789***                                                 & -0.440***                                                 \\
				$Wlny$                      & -                                                                            & 0.528***                                                  & 0.545***                                                  & 0.732***                                                  & 0.749***                                                  & 0.516***                                                  & 0.505***                                                  & 0.725***                                                  & 0.726***                                                  \\
				Moran's I                  & \begin{tabular}[c]{@{}c@{}}0.360*** ($W^d_{1n}$)\\      0.431*** ($W^d_{2n}$)\end{tabular} & -                                                         &                                                           & -                                                         & -                                                         & -                                                         & -                                                         & -                                                         & -                                                         \\
				Test of Restriction      & \begin{tabular}[c]{@{}c@{}}10.944***\\      (Wald)\end{tabular}              & \begin{tabular}[c]{@{}c@{}}2.742\\      (LR)\end{tabular} & \begin{tabular}[c]{@{}c@{}}2.688\\      (LR)\end{tabular} & \begin{tabular}[c]{@{}c@{}}0.384\\      (LR)\end{tabular} & \begin{tabular}[c]{@{}c@{}}3.458\\      (LR)\end{tabular} & \begin{tabular}[c]{@{}c@{}}2.351\\      (LR)\end{tabular} & \begin{tabular}[c]{@{}c@{}}2.363\\      (LR)\end{tabular} & \begin{tabular}[c]{@{}c@{}}4.074\\      (LR)\end{tabular} & \begin{tabular}[c]{@{}c@{}}0.196\\      (LR)\end{tabular} \\
				$\alpha$                      & 0.697***                                                                     & 2.516                                                     & 2.737                                                     & -0.134                                                    & -0.184                                                    & -2.855                                                    & -3.703                                                    & -0.901                                                    & -0.926                                                    \\
				$\phi$                        & -                                                                            & -1.946                                                    & -2.169                                                    & 0.479**                                                   & 0.556*                                                    & 3.431                                                     & 4.280                                                     & 1.206\&                                                   & 1.307                                                     \\
				$\gamma$                      & -                                                                            & -0.150                                                    & -0.134                                                    & 0.458***                                                  & 0.411***                                                  & 0.056                                                     & 0.045                                                     & 0.274**                                                   & 0.235\&                                                   \\
				$\alpha+\phi/(1-\gamma)$                         & -                                                                            & 0.823***                                                  & 0.824***                                                  & 0.749***                                                  & 0.759***                                                  & 0.780***                                                  & 0.780***                                                  & 0.759***                                                  & 0.783***                                                  \\
				Presence of $\phi$ (Wald)                 & -                                                                            & 2.000                                                     & 1.592                                                     & 7.507                                                     & 4.023                                                     & 1.332                                                     & 1.016                                                     & 10.675                                                    & 5.784                                                     \\
				\hline
				\multicolumn{10}{r}{(Significant at *: 1\%, **: 5\%, ***: 10\%, \&: 15\%; $s*$ is a tuning parameter and set as 2.5 here.)} \\                            
			\end{tabular}
		\end{minipage}
	\end{sidewaystable}

	\begin{sidewaystable}
		\caption{Results using Penn World Table verison 6.1 (Year 1960-1995)}
		\begin{minipage}{0.5\textwidth}
			\tiny
			\begin{tabular}{c|c|cccc|cccc} \cline{2-10}
				&                                                                              & & \multicolumn{3}{c}{$W^d_{1n} \circ W^e_n$}  \vline                                                                                                                                                                                                                     & \multicolumn{4}{c}{$W^d_{2n} \circ W^e_n$}                                                                                                                                                                                                                       \\
				& Solow-Swan                                                                        &  \multicolumn{1}{c}{$W^d_{1n}$}                                                   & $exp(-2d_{ij})$                                                       & $\frac{1}{(|z_i-z_j|^{2 \cdot s*})}$                                    & $\frac{1}{(min\left(\frac{z_i}{z_j},\frac{z_j}{z_i}\right)-1)^2}$                                                     &        $W^d_{2n}$                                               & $exp(-2d_{ij})$                                                      & $\frac{1}{(|z_i-z_j|^{2 \cdot s*})}$                                       & $\frac{1}{(min\left(\frac{z_i}{z_j},\frac{z_j}{z_i}\right)-1)^2}$                                                     \\ \hline
				Model & (1) & (2) & (3) & (4) & (5) & (6) & (7) & (8) & (9) \\  \hline
				Endogeneity test (RS)  & -                                                                            & -                                                         & 28.527***                                                 & 14.150***                                                 & 9.074***                                                  & -                                                         & 25.337***                                                 & 17.720***                                                 & 17.696***                                                  \\
				Constant terms                      & 4.651**                                                                      & 0.988                                                     & 1.159***                                                  & 4.189***                                                  & 5.959***                                                  & 0.530                                                     & 1.386***                                                  & 4.441***                                                  & 6.236***                                                   \\
				$lns$                        & 1.276***                                                                     & 0.825***                                                  & 0.811                                                     & 0.596                                                     & 0.538                                                     & 0.792***                                                  & 0.777                                                     & 0.499                                                     & 0.499                                                      \\
				$ln(n+g+\delta)$                      & -2.709***                                                                    & -1.498***                                                 & -1.459***                                                 & -0.976***                                                 & -0.760***                                                 & -1.451***                                                 & -1.452***                                                 & -1.149***                                                 & -0.444***                                                  \\
				$Wlns$                       & -                                                                            & -0.322**                                                  & -0.287***                                                 & 0.316***                                                  & 0.420***                                                  & -0.372**                                                  & -0.310***                                                 & 0.392***                                                  & 0.425***                                                   \\
				$Wln(n+g+\delta)$                     & -                                                                            & 0.571                                                     & 0.565***                                                  & 0.084***                                                  & 0.769***                                                  & 0.137                                                     & 0.423***                                                  & 0.546***                                                  & 0.626***                                                   \\
				$Wlny$                      & -                                                                            & 0.740***                                                  & 0.736***                                                  & 0.494***                                                  & 0.573***                                                  & 0.658***                                                  & 0.657***                                                  & 0.543***                                                  & 0.585***                                                   \\
				Moran's I                  & \begin{tabular}[c]{@{}c@{}}0.410*** (Wd2)\\      0.436*** (We2)\end{tabular} & -                                                         & -                                                         & -                                                         & -                                                         & -                                                         & -                                                         & -                                                         & -                                                          \\
				Test of Restriction      & \begin{tabular}[c]{@{}c@{}}4.427**\\      (Wald)\end{tabular}                & \begin{tabular}[c]{@{}c@{}}1.576\\      (LR)\end{tabular} & \begin{tabular}[c]{@{}c@{}}1.458\\      (LR)\end{tabular} & \begin{tabular}[c]{@{}c@{}}0.767\\      (LR)\end{tabular} & \begin{tabular}[c]{@{}c@{}}4.298\\      (LR)\end{tabular} & \begin{tabular}[c]{@{}c@{}}2.338\\      (LR)\end{tabular} & \begin{tabular}[c]{@{}c@{}}1.962\\      (LR)\end{tabular} & \begin{tabular}[c]{@{}c@{}}4.409\\      (LR)\end{tabular} & \begin{tabular}[c]{@{}c@{}}5.346*\\      (LR)\end{tabular} \\
				$\alpha $                     & 0.580***                                                                     & 0.276**                                                   & 0.252**                                                   & -1.782                                                    & -1.252                                                    & 0.299**                                                   & 0.275**                                                   & -2.379                                                    & -                                                          \\
				$\phi$                        & -                                                                            & 0.180*                                                    & 0.200*                                                    & 2.147                                                     & 1.593                                                     & 0.151\&                                                   & 0.167\&                                                   & 2.710                                                     & -                                                          \\
				$\gamma$                      & -                                                                            & 0.557***                                                  & 0.536***                                                  & 0.113                                                     & 0.164                                                     & 0.508***                                                  & 0.499***                                                  & 0.105                                                     & -                                                          \\
				$\alpha+\phi/(1-\gamma)$                         & -                                                                            & 0.683***                                                  & 0.684***                                                  & 0.639***                                                  & 0.654***                                                  & 0.606***                                                  & 0.609***                                                  & 0.650                                                     & -                                                          \\
				Presence of $\phi$ (Wald)                 & -                                                                            & 1.463                                                     & 1.330                                                     & 10.802***                                                 & -0.628                                                    & 1.059                                                     & 0.921                                                     & 14.180                                                    & -                                                          \\
				\hline
				\multicolumn{10}{r}{(Significant at *: 1\%, **: 5\%, ***: 10\%, \&: 15\%; $s*$ is a tuning parameter and set as 2.5 here.)}                                                
			\end{tabular}
		\end{minipage}
	\end{sidewaystable}
	
	\section{Conclusion}
	
	\noindent Growth theory has been one of the mainly important issues in economics. Questions are, for example, about the source of the gap between poor and rich countries, or whether the gap would converge to the equilibrium or not. Shedding light on the link from the geographical location to economic growth, Ertur and Koch (2007) found a significant evidence that the spatial information works effectively to explain the growth pattern in the economy. More specifically, Ertur and Koch (2007) built up the growth model where the aggregate level of technology was constructed through home externalities pertaining to knowledge as well as physical/geographical spatial externalities or technological interdependence across countries. The spatial information, however, was solely based on the physical great-circle distance and hence depended on the exogenous assumption for the spatial weights matrices ($W$). However, empricial literature find supportive evidence that economic interactions such as international trade volume distribute the technology/knowledge diffusion. In this spirit, I investigated how the results would become different if the exogenous $W$ assumption is violated or more generalized to include economic distance. In Qu and Lee (2015), $W$ is allowed to be a function of a set of random variables and thus \textit{stochastic}. This leads to endogenous $W$ unless the covariance of disturbance terms in the auxiliary and final outcome equations are zero. \\
	
	The main dataset is from Penn World Table version 7.1 over the year 1960-2010, which includes fifteen more years (1996-2010) than in Ertur and Koch (2007). Note that those years experienced the worldwide financial crisis in late 1990s and 2000s and physical capital has fluctuated more than ever for that cause. Three types of economic distance were considered. Taking GDI as a key constructor for the economic distance, I considered the form of $W$ being Hadamard product of the physical distance and economic distance as in Qu and Lee (2015). I then conducted the robust Rao's score test (Bera, Dogan, and Taspinar, 2018) to determine if $W$ is exogenous. The results rejected the null and one may conclude that the $W$'s are endogenous. I then applied the ML estimation (Qu and Lee, 2015) and found a couple of interesting results that the spatially augmented Solow-Swan model with the exogenous $W$ gives conflicting results with those in Ertur and Koch (2007). Neither any of the parameters (Physical capital share $\alpha$; Physical capital externalities $\phi$; Technological interdependence $\gamma$) were significant nor the signs $\phi, \gamma$ were consistent but opposite, depending on the physical weights matrices. The presence of $\phi$ was not significant, either. But the models with the endogenous $W$ constructed by economic distance could estimate significant and positive effects of $\phi$ \& $\gamma$ as well as strong significance in presence of $\phi$. All spatially-augmented models did not find $\alpha$ significant and its values were out of our expectation, which might be from the fact that financial crisis made the effect of $\alpha$ unclear. \\
	
	I also found an empirical strategy that the power distance weight function type worked better in estimating $\phi$ and $\gamma$ if the data cover recent years under high fluctuations, wherefore the shocks have not been resolved yet. In this case, $W$ is scale-sensitive with respect to the physical weights matrices ($W^d$) since $W^d$ affects the path along which the economic distance works on spatial interactions. This may lead to substantial difference in estimates, although the signs and significances are consistent. Hence one needs to decide an appropriate $W^d$ carefully if the interpretation in magnitude of $\phi$ and $\gamma$ is important. If the period covered in data is in relatively stable status or has had enough time to get back to a stable equilibrium, the negative exponential function type performed nicely. Interestingly, the results were almost identical to those with the exogenous $W$. \\
	
	The results in this study imply that exogenous $W$ is not robust to capture the spatial intercation effects if some abnormal shocks occur since the physical distance is invariant but not affected by the change in economy. Economic variables, however, are closely synchronized to a shock and therefore have inherent information regarding the economic shocks: For example, Zhou (1995) analyzed the response of the real exchange rates to various types of economic shocks; Evans and Marshall (2009) studied how macroeconomic/financial variables respond to shocks in various sectors such as technology, labor supply, and monetary policy; Shields, Olekalns, Henry, and Brooks (2005) found asymmetric response of inflation and growth volatility to shocks. Findings in this study are coherent with the existing empirical literature emphasizing the importance of economic distance. Considering economic distance as a part of the spatial weights matrices with the endogeneity of $W$ controlled appropriately hence benefits us in that one can deal with the aftermath from some unexpected shocks on the economic dependent variable, i.e., one may expect it to be \textit{robust} to uncover the underlying spatial interaction effects.

	\newpage
	\section*{Reference}
	\scriptsize
	
	
	Barro, Robert. (1991).  Economic Growth in a Cross Section of Countries. \textit{The Quarterly Journal of Economics}, 106(2): 407-443.
	
	
	
	Baxter, M., Kouparitsas, M. (2005). Determinants of business cycle comovement: a robust analysis. \textit{Journal of Monetary Economics}, 52: 113–157.
	
	
	Bera, Anil K., Dogan, Osman., and Taspinar, Suleyman. (2018). Simple tests for endogeneity of spatial weights matrices. \textit{Regional Science and Urban Economics}, 69: 130-142.
	
	
	
	
	
	Chen, Yanguang. (2015). The distance-decay function of geographical gravity model: Power law or exponential law?. \textit{Chaos, Solitions and Fractals}, 77: 174-189.
	
	Cheng, Wei., Lee, Lung Fei. (2017). Testing endogeneity of spatial and social networks. \textit{Regional Science and Urban Economics}, 64: 81-97.
	
	
	
	De Long, J. Bradford., Summers, Lawrence H. (1991). Equipment Investment and Economic Growth, \textit{The Quarterly Journal of Economics}. 106(2): 445–502.
	
	
	Ditzen, Jan. (2018). Cross-country convergence in a general Lotka–Volterra model. \textit{Spatial Economic Analysis}, 13(2): 191-211.
	
	Domar, Evsey (1946). Capital Expansion, Rate of Growth, and Employment. \textit{Econometrica}, 14(2): 137–147.
	
	Durbin, J. (1960). Estimation of Parameters in TimeSeries Regression Models. \textit{Journal of the Royal Statistical Sociely},  22(1): 139-53. 
	
	
	Ertur, Cem. and Koch, Wilfried. (2007). Growth, technological interdependence and spatial externalities: theory and evidence. \textit{Journal of Applied Econometrics}, 22(6): 1033-1062.
	
	Ertur, Cem. and Koch, Wilfried. (2011). A contribution to the theory and empirics of Schumpeterian growth with worldwide interactions. \textit{Journal of Economic Growth}, 16(3): 215-255.
	
	Evans, Charles L., Marshall, David A. (2009). Fundamental Economic Shocks and the Macroeconomy. \textit{Journal of Money, Credit and Banking}, 41(8): 1515-1555.
	
	
	
	
	Frankel, J., Rose, A. (1998). The endogeneity of the optimum currency area criteria. \textit{The Economic Journal}, 108: 1009–1025.
	
	
	
	
	Harrod, Roy F. (1939). An Essay in Dynamic Theory. \textit{The Economic Journal}, 49(193): 14–33.
	
	Ho, Chun-Yu., Wang, Wei., Yu, Jihai. (2013). Growth spillover through trade: A spatial dynamic panel data approach. \textit{Economics Letters}, 120(3):450–453.
	
	
	
	Jenish, Nazgul., Prucha, Ingmar R. (2009). Central limit theorems and uniform laws of large numbers for arrays of random fields. \textit{Journal of Econometrics}, 150(1): 86-98.
	
	Jenish, Nazgul., Prucha, Ingmar R. (2012). On spatial processes and asymptotic inference under near-epoch dependence. \textit{Journal of Econometrics}, 170(1): 178-190.
	
	Keller, Wolfgang. (2002). Geographic Localization of International Technology Diffusion. \textit{American Economic Review}, 92(1): 120-142.
	
	Klenow, Peter J., Rodriguez-Clare, Andres. (2005). Externalities and Growth. \textit{Handbook of Economic Growth}, in: Philippe.
	
	
	
	
	
	Lucas, Robert Jr. (1988). On the mechanics of economic development. \textit{Journal of Monetary Economics}, 22(1): 3-42.
	
	Lucas, Robert. (1993). Making a Miracle. \textit{Econometrica}, 61(2): 251-72.
	
	Mankiw, N. Gregory., Romer, David., Weil, David N. (1992). A Contribution to the Empirics of Economic Growth. \textit{The Quarterly Journal of Economics}, 107(2): 407–437. 
	
	Meade, James. (1961). \textit{A Neo-classical Theory of Economic Growth}. London: Unwin.
	
	
	
	
	
	
	Nalewaik, Jeremy J. (2012). Estimating Probabilities of Recession in Real Time Using GDP and GDI. \textit{Journal of Money, Credit and Banking}, 44(1): 235-253.
	
	Qu, Xi., and Lee, Lung-fei. (2015). Estimating a spatial autoregressive model with an endogenous spatial weight matrix. \textit{Journal of Econometrics}, 184(2): 209-232.
	
	
	
	Romer, Paul M. (1989). Human Capital And Growth: Theory and Evidence. \textit{NBER Working Paper, No.3173, NBER Program(s): The Economic Fluctuations and Growth Program}.
	
	
	
	
	Shields, Kalvinder., Olekalns, Nilss., Henry, Ólan T., Brooks, Chris. (2005).  Measuring the Response of Macroeconomic Uncertainty to Shocks. \textit{The Review of Economics and Statistics}, 87(2): 362–370.
	
	Solow, Robert M. (1956). A Contribution to the Theory of Economic Growth. \textit{The Quarterly Journal of Economics}, 70(1): 65-94.
	
	Swan, Trevor W. (1956). Economic growth and capital accumulation. \textit{Economic Record}, 32(2): 334–361.
	
	
	Temple, Jonathan. (1999). The New Growth Evidence. \textit{Journal of Economic Literature}, 37(1): 112-156.
	
	
	
	Zhou, Su. (1995). The Response of Real Exchange Rates to Various Economic Shocks. \textit{Southern Economic Journal}, 61(4): 936-954.
	
\end{document}